\documentclass[useAMS,usenatbib]{mn2e}

\usepackage{graphicx}
\usepackage{epstopdf}

\title[Spherical Needlets for CMB Data Analysis]{Spherical Needlets for CMB Data Analysis}
\author[D. Marinucci et al.]{D.~Marinucci$^1$\thanks{E-mail: marinucc@mat.uniroma2.it}, D.~Pietrobon$^2$, A.~Balbi$^{2,3}$, P.~Baldi$^{1}$, P.~Cabella$^4$, \newauthor G.~Kerkyacharian$^5$,
 P.~Natoli$^{2,3}$, D.~Picard$^6$, N.~Vittorio$^{2,3}$\\
\\
$1$ Dipartimento di Matematica, Universit\`a di Roma ``Tor Vergata'', Via della Ricerca Scientifica 1, 00133 Roma \\
$2$ Dipartimento di Fisica, Universit\`a di Roma ``Tor Vergata'', Via della Ricerca Scientifica 1, 00133 Roma \\
$3$ INFN Sezione di Roma ``Tor Vergata'', Via della Ricerca Scientifica 1, 00133 Roma \\
$4$ University of Oxford, Astrophysics, Keble Road, Oxford, OX1 3RH, U.K.\\
$5$ Universit\'e de Paris 10 and Laboratoire de Probabilit\'es et Mod\`eles Al\'eatoires\\
$6$ Universit\'e de Paris 7 and Laboratoire de Probabilit\'es et Mod\`eles Al\'eatoires}

\date{\today}

\begin{document}

\maketitle

\begin{abstract}
We discuss Spherical Needlets and their properties. Needlets are a form of
spherical wavelets which do not rely on any kind of tangent plane
approximation and enjoy good localization properties in both pixel and harmonic space; moreover needlets coefficients are asymptotically
uncorrelated at any fixed angular distance, which makes their use in
statistical procedures very promising. In view of these properties, we
believe needlets may turn out to be especially useful in the analysis of
Cosmic Microwave Background (CMB) data on the incomplete sky, as well as of other cosmological observations. As a final advantage, we
stress that the implementation of needlets is computationally very
convenient and may rely completely on standard data analysis packages such as HEALPix.
\end{abstract}





\section{Introduction}\label{introduction}

Over the last few years, wavelets have emerged as one of the most powerful
tools of CMB data analysis, finding applications in virtually all areas
where statistical methods are required; a very incomplete list of references
should include testing for non-Gaussianity (see \citet{Vielva2004,Cabella2004}), foreground subtraction (\citet{Hansen2006}), point source detection (\citet{Sanz2006}), component separation (\citet{Moudden2005,Starck2005}), polarization analysis (\citet{CabellaNatoliSilk2007}) and many others. The reason for such a
strong interest is easily understood. As it is well-known, CMB models are best
analyzed in the frequency domain, where the behaviour at different
multipoles can be investigated separately; on the other hand, partial sky
coverage and other missing observations make the evaluation of
exact spherical harmonic transforms troublesome. The combination of these two
features makes the time-frequency localization properties of wavelets most
valuable.

Despite the wide agreement on their importance as a data analysis instrument, the derivation of an optimal wavelets basis on the sphere is still an open issue for research. Many efforts have been undertaken in this area, most of them being based upon the so-called tangent plane approach \citep{MR1721807}. In this framework,  a flat sky
approximation is entertained locally, and then some form of standard plane wavelets are
implemented. Directional wavelets
have been advocated by \cite{McEwen2006, McEwen2007}, again by means of a tangent
plane approximation. An interesting attempt to overcome the tangent plane approximation
is due to \cite{Sanz2006}.

A new approach to spherical wavelets was introduced in the statistical
literature by \cite{Baldi2006}, adapting tools proposed in the
functional analysis literature by \cite{NarcowichPetrushevWard2006}; the first application to CMB data is due to \cite{PietrobonBalbiMarinucci2006}, (see also \cite{Cardoso2007,2007arXiv0706.4169B}). The idea is to focus on so-called needlets, to be described in the
following section. Needlets enjoy several features which are not in general
granted by other spherical wavelets construction; we anticipate some of
these features, which we shall investigate more deeply in the Sections to
come. More precisely:

a) they do not rely on any tangent plane approximation (compare \citealt{Sanz2006}), and take advantage of the manifold structure of the sphere;

b) being defined in harmonic space, they are computationally very convenient, and natively adapted to
standard packages such as HEALPix\footnote{http://healpix.jpl.nasa.gov} \citep{2005ApJ...622..759G};

c) they allow for a simple reconstruction formula (see Eq.~\ref{recfor}),
where the same needlets functions appear both in the direct and the inverse
transform (see also \citealt{Kerk2007}). This property is the same as for spherical harmonics but it is 
\emph{not }shared by other wavelets systems such as the well-known Spherical
Mexican Hat Wavelets (hereafter SMHW);

d) they are quasi-exponentially (i.e. faster than any polynomial) concentrated in pixel space, see Eq.~\ref{expine} below;

e) they are exactly localized on a finite number of multipoles; the
width of this support is explicitly known and can be specified as an input parameter (see Eq.~\ref{needlets_expansion});

f) random needlets coefficients can be shown to be asymptotically uncorrelated
(and hence, in the Gaussian case, independent) at any fixed angular
distance, when the frequency increases. This capital property can be
exploited in several statistical procedures, as it allows to treat needlets
coefficients as a sample of independent and identically distributed
coefficients on small scales, at least under the Gaussianity assumption.

The aim of this paper is to discuss more thoroughly the implementation of needlets, compare it with other wavelets (namely, the SMHW)  and investigate
their properties by means of Monte Carlo simulations. In Section \ref{implementation} we
describe the numerical implementation, taking care to discuss the features
of the construction that ensure the above-mentioned properties. In Section \ref{properties}
we discuss the relationship between the localization properties in frequency
and pixel spaces; we also discuss the trade-off between the two, which from
the operational point of view relates to the issue of an optimal choice of
a so-called ``bandwidth'' parameter $B$. Some
comparisons are made with existing techniques to deal with partial sky coverage, and
more precisely with tophat binning procedures and SMHW. Some
discussion on possible applications and directions for further research is
collected in Section \ref{conclusions}.

\section{The Numerical Implementation of Needlets}\label{implementation}

We start  by outlining briefly the construction of a needlets basis.
More details can be found in \cite{NarcowichPetrushevWard2006}, and in \cite{Baldi2006}. We shall discuss the
details of the construction step by step, in order to provide to
potential users a clear recipe for needlets implementation. 

We first recall that the spherical needlet (function) is defined as 
\begin{equation}
\psi _{jk}(\hat\gamma) =\sqrt{\lambda_{jk}}\sum_{\ell}b(\frac{\ell}{B^{j}})\sum_{m=-\ell}^{\ell}\overline{Y}%
_{\ell m}(\hat\gamma)Y_{\ell m}(\xi _{jk});  \label{needlets_expansion}
\end{equation}%
here, we use $\left\{ \xi _{jk}\right\} $ to denote a set of \emph{cubature
points }on the sphere, corresponding to frequency $j;$ in practice, we shall
identify these points with the pixel centres in HEALPix. Also, $\lambda_{jk}$ denotes the cubature weights, which for simplicity can be envisaged as $1/N_{p}$, $N_p$ denoting the number of pixels (see \citealt{PietrobonBalbiMarinucci2006}).

\begin{figure}
\includegraphics[width=\columnwidth]{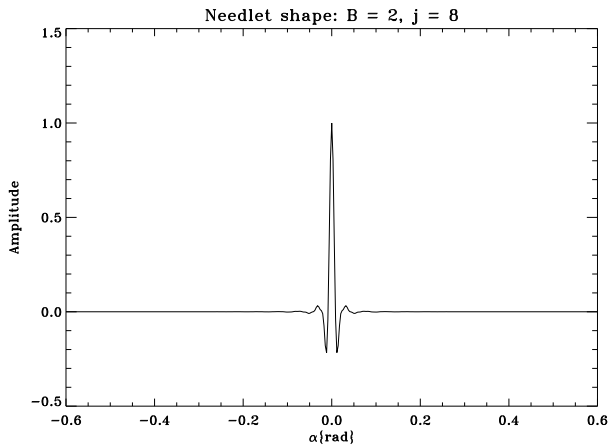}
\caption[]{Needlets in pixel space. $B=2$, $j=8$}
\end{figure}

Intuitively, needlets should be viewed as a convolution of the projection
operator $\sum_{m=-\ell}^{\ell}\overline{Y}%
_{\ell m}(\hat\gamma)Y_{\ell m}(\xi _{jk})$  with a suitably chosen window function 
$b(\cdot)$ Special properties of $b(\cdot)$ ensure that the needlets enjoy
quasi-exponential localization properties in pixel space. Formally, we must ensure
that \citep{NarcowichPetrushevWard2006,Baldi2006}:

\begin{enumerate}
\item $b^{2}(\cdot)$ has support in $[\frac{1}{B},B],$ and hence $b(\frac{\ell}{%
B^{j}})$ has support in $\ell\in \lbrack B^{j-1},B^{j+1}]$

\item the function $b(\cdot)$ is infinitely differentiable in $(0,\infty ).$

\item we have%
\begin{equation}
\sum_{j=1}^{\infty }b^{2}(\frac{\ell}{B^{j}})\equiv 1\textrm{ for all }\ell>B.%
  \label{partun}
\end{equation}
\end{enumerate}

It is immediate to see that property (i) ensures the needlets have bounded
support in the harmonic domain; property (ii) is the crucial element in the
derivation of the localization properties, which we shall illustrate in the
following section. Finally, property (iii) is necessary to establish the
reconstruction formula which we shall discuss below; functions such as $b^{2}(\cdot)$ are called \emph{partitions of
unity}.

There are of course many possible constructions satisfying
(i-iii); indeed an interesting theme for future research is the derivation
of optimal windows satisfying these three conditions (compare \citealt{Cardoso2007}). An explicit recipe for the
construction of  $b(\cdot)$ is as follows.

\begin{enumerate}
\item[STEP 1:] Construct the function
\[
f(t)=\left\{ 
\begin{array}{ll}
\exp (-\frac{1}{1-t^{2}})\textrm{ , } & -1\leq t\leq 1 \\ 
0, & \textrm{otherwise }
\end{array}
\right. .
\]
It is immediate to check that the function $f(\cdot)$ is $C^{\infty }$ and
compactly supported in the interval $(-1,1)$

\item[STEP 2:] Construct the function 
\[
\psi (u)=\frac{\int_{-1}^{u}f(t)dt}{\int_{-1}^{1}f(t)dt}.
\]%
The function $\psi (\cdot)$ is again $C^{\infty };$ it is moreover
non-decreasing and normalized so that $\psi (-1)=0$ , $\psi (1)=1$

\item[STEP 3:] Construct the function%
\[
\varphi (t)=\left\{ 
\begin{array}{lllll}
1 & \textrm{ if } & 0\leq & t & \leq \frac{1}{B} \\ 
\psi (1-\frac{2B}{B-1}(t-\frac{1}{B})) & \textrm{ if } & \frac{1}{B}\leq & t & \leq 1 \\ 
0 & \textrm{ if } & & t & >1
\end{array}
\right. 
\]
Here we are simply implementing a change of variable so that the resulting
function $\varphi (\cdot)$ is constant on $(0,B^{-1})$ and monotonically
decreasing to zero in the interval $(B^{-1},1).$ Indeed it can be checked
that%
\[
1-\frac{2B}{B-1}(t-\frac{1}{B})=\left\{ 
\begin{array}{ccc}
1 & \textrm{ for } & t=\frac{1}{B} \\ 
-1 & \textrm{ for } & t=1
\end{array}
\right. 
\]
and
\begin{eqnarray*}
\varphi (\frac{1}{B}) &=&\psi (1)=1 \\
\varphi (1) &=&\psi (-1)=0
\end{eqnarray*}

\item[STEP 4:] Construct 
\[
b^{2}(\xi )=\varphi (\frac{\xi }{B})-\varphi (\xi )
\]%
The expression for $b^{2}(\cdot)$ is meant
to ensure that the function satisfies the partition-of-unity property of Eq.~\ref
{partun}. Needless to say, for  $b(\xi )=\left\{ \varphi (\frac{\xi }{B}%
)-\varphi (\xi )\right\} ^{1/2}$ we take the positive root.

\ 
\end{enumerate}

Random needlets coefficients are hence given by%
\begin{eqnarray}
\beta _{jk} &=&\int_{S^{2}}T(\hat\gamma)\psi _{jk}(\hat\gamma)d\Omega  \nonumber \\
&=&\sqrt{\lambda_{jk}}\sum_{\ell}b(\frac{\ell}{B^{j}})\sum_{m=-\ell}^{\ell}\left\{
\int_{S^{2}}T(\hat\gamma)\overline{Y}_{\ell m}(\hat\gamma)d\Omega\right\} Y_{\ell m}(\xi _{jk})  \nonumber
\\
&=&\sqrt{\lambda_{jk}}\sum_{\ell}b(\frac{\ell}{B^{j}})\sum_{m=-\ell}^{\ell}a_{\ell m}Y_{\ell m}(%
\xi _{jk}).  \label{needcof}
\end{eqnarray}

It is very important to stress that, although the needlets do \emph{not} make
up an orthonormal basis for square integrable functions on the sphere, they
do represent a \emph{tight frame. }In general, a tight frame on the sphere
is a countable set of functions $\left\{ e_{j}\right\} $ such that, for all
square integrable functions on the sphere $f\in L^{2}(S^{2}),$ we have%
\[
\sum_{j}\langle f,e_{j}\rangle ^2\equiv \int_{S^{2}}f(\hat\gamma)^{2}d\Omega,
\]
so that the norm is preserved. Of course, this norm-preserving property is
shared by all orthonormal systems; however, frames do not in general make up
a basis, as they admit redundant elements. They can be viewed as the closer
system to a basis, for a given redundancy, see \cite{HernandezWeiss1996}, \cite{Baldi2006} for further definitions and discussion. In our framework, the
norm-preserving property becomes  
\begin{equation}
\sum_{j,k}\hat\beta _{jk}^{2}\equiv \sum_{\ell=1}^{\infty }\frac{2\ell+1}{4\pi }%
\widehat{C}_{\ell}\textrm{ ,}  \label{frame}
\end{equation}
where 
\[
\widehat{C}_\ell= \frac{1}{2\ell+1}\sum_m |a_{\ell m}|^2
\]
is the raw angular power spectrum of the map $T(\hat\gamma)$. The identity in Eq. \ref{frame} has indeed been verified by means of numerical simulations and implicitly provides the correct normalization for needlets. It is basically a consequence of the peculiar partition-of-unity property of $b(\cdot)$ of Eq. \ref{partun}. Of course this property is not generally shared by other constructions such as SMHW, where the wavelets functions are normalized to unity in the real domain. Eq.~\ref{frame} is related to a much more fundamental result, i.e.
the reconstruction formula 
\begin{equation}
T(\hat\gamma)\equiv \sum_{j,k}\beta _{jk}\psi _{jk}(\hat\gamma)
\label{recfor}
\end{equation}%
which in turn is a non-trivial consequence of the careful construction
leading to Eq.~\ref{partun}. As mentioned before, the simple reconstruction formula of Eq.~\ref{recfor} is typical of tight frames but does not hold in general for other wavelets systems. It
is easy to envisage many possible applications of this result
when handling masked data.

\section{Properties and comparisons}\label{properties}

The following quasi-exponential localization property of needlets is due to
\cite{NarcowichPetrushevWard2006} and motivates their name.

For any $k=1,2,...$ there exists a positive constant $c_{k}$ such that for
any point $\hat\gamma\in S^{2}$ we have%
\begin{equation}
|\psi _{jk}(\hat\gamma)|\leq \frac{c_{k}B^{j}}{(1+B^{j}\arccos (\langle\hat\gamma,\xi _{jk}\rangle)^{k}}%
\textrm{ .}  \label{expine}
\end{equation}

We recall that $\arccos (\langle\hat\gamma,\xi _{jk}\rangle)$ is just the natural distance on the
unit sphere between the points $(\hat\gamma,\xi _{jk}).$ The meaning of Eq.~\ref{expine} is then clear: for any fixed angular distance, the value of $\psi _{jk}(\hat\gamma)$
goes to zero quasi-exponentially in the parameter $B.$ The resulting trade-off in
the behaviour over the harmonic and real spaces is expected: smaller values
of $B$ correspond to a tighter localization in harmonic space (less
multipoles entering into any needlet), whereas larger values ensure a faster
decay in real space.

\begin{figure}
\begin{center}
\includegraphics[angle=90, width=\columnwidth]{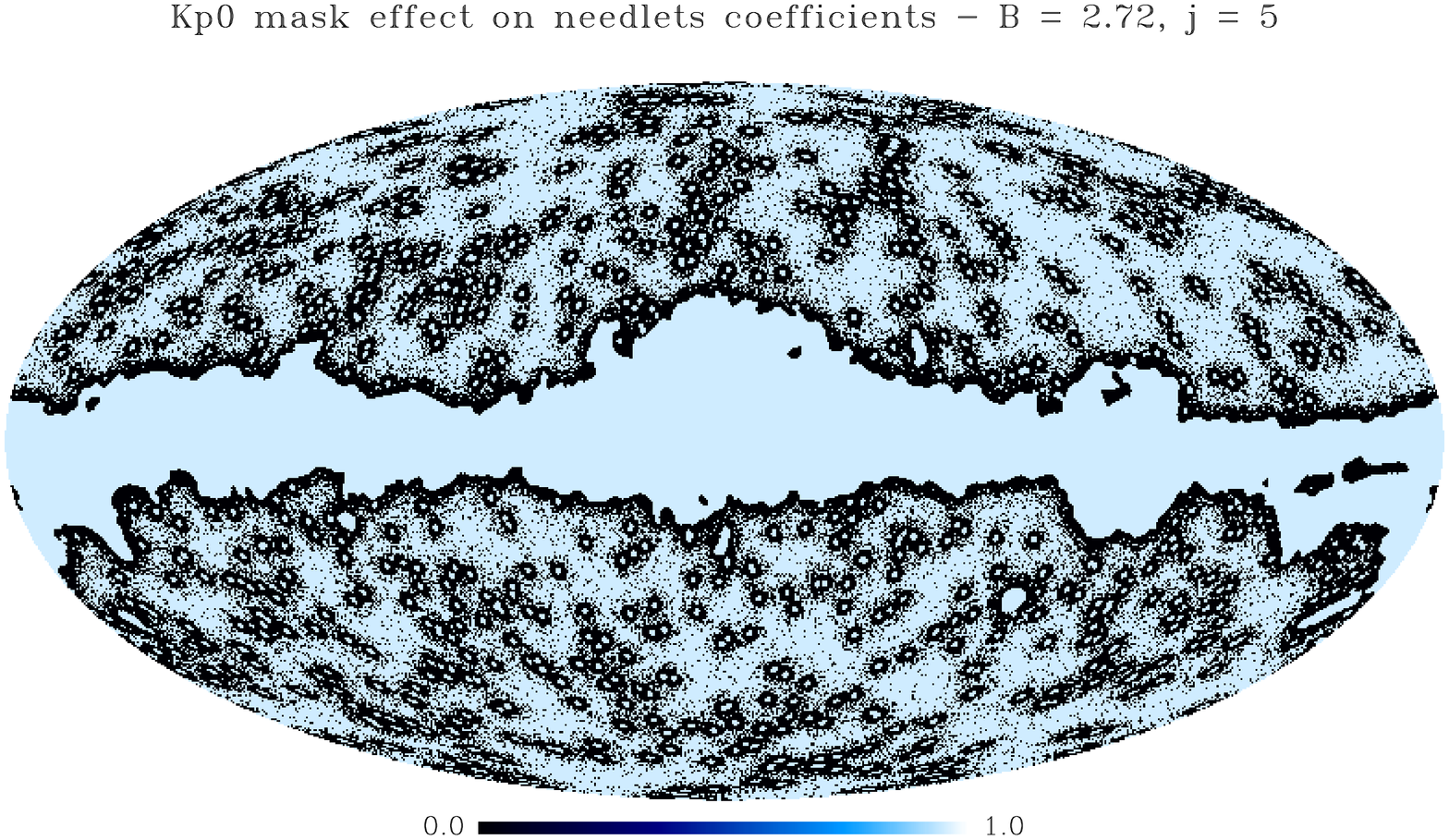}
\includegraphics[angle=90, width=\columnwidth]{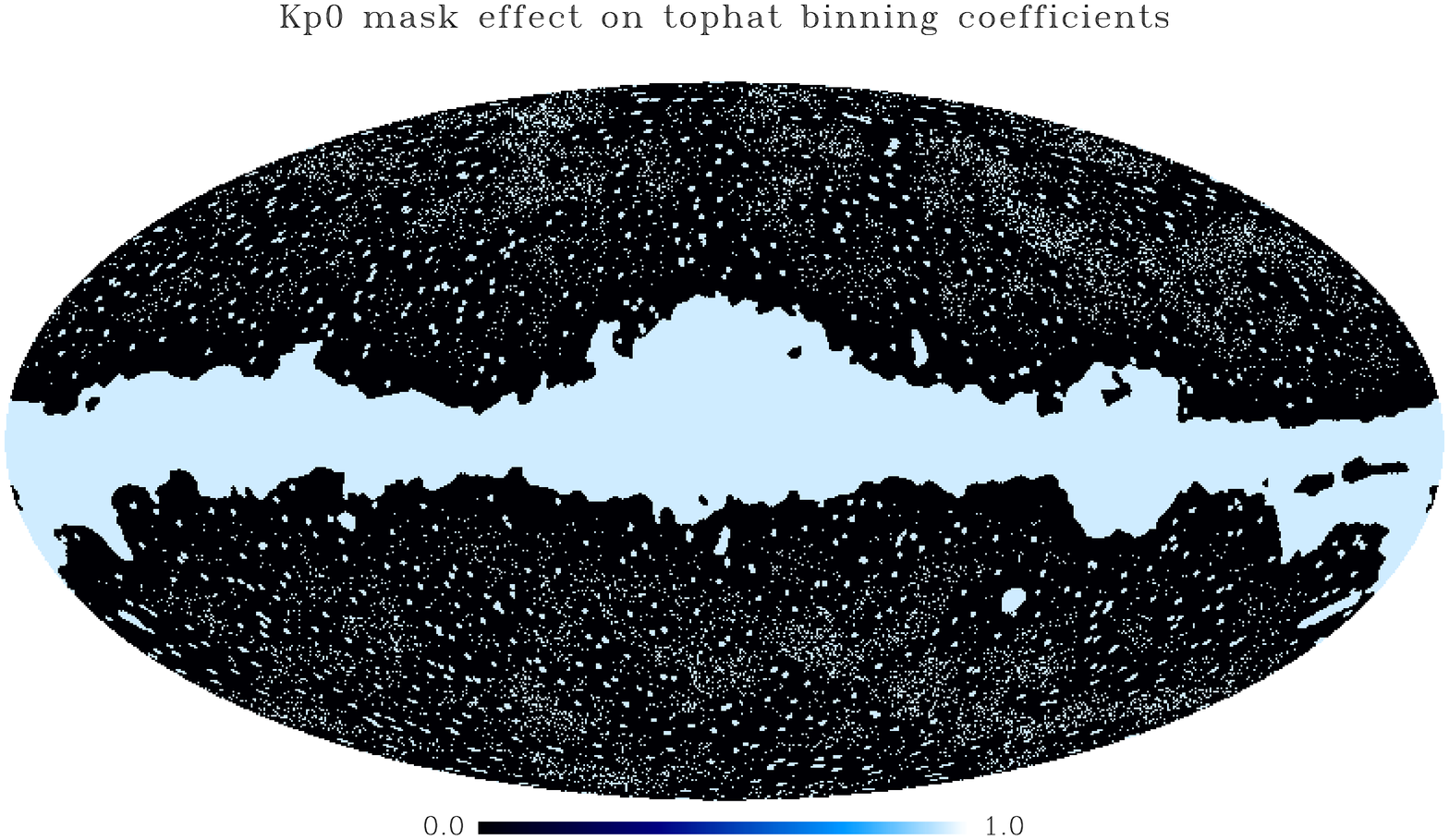}
\includegraphics[angle=90, width=\columnwidth]{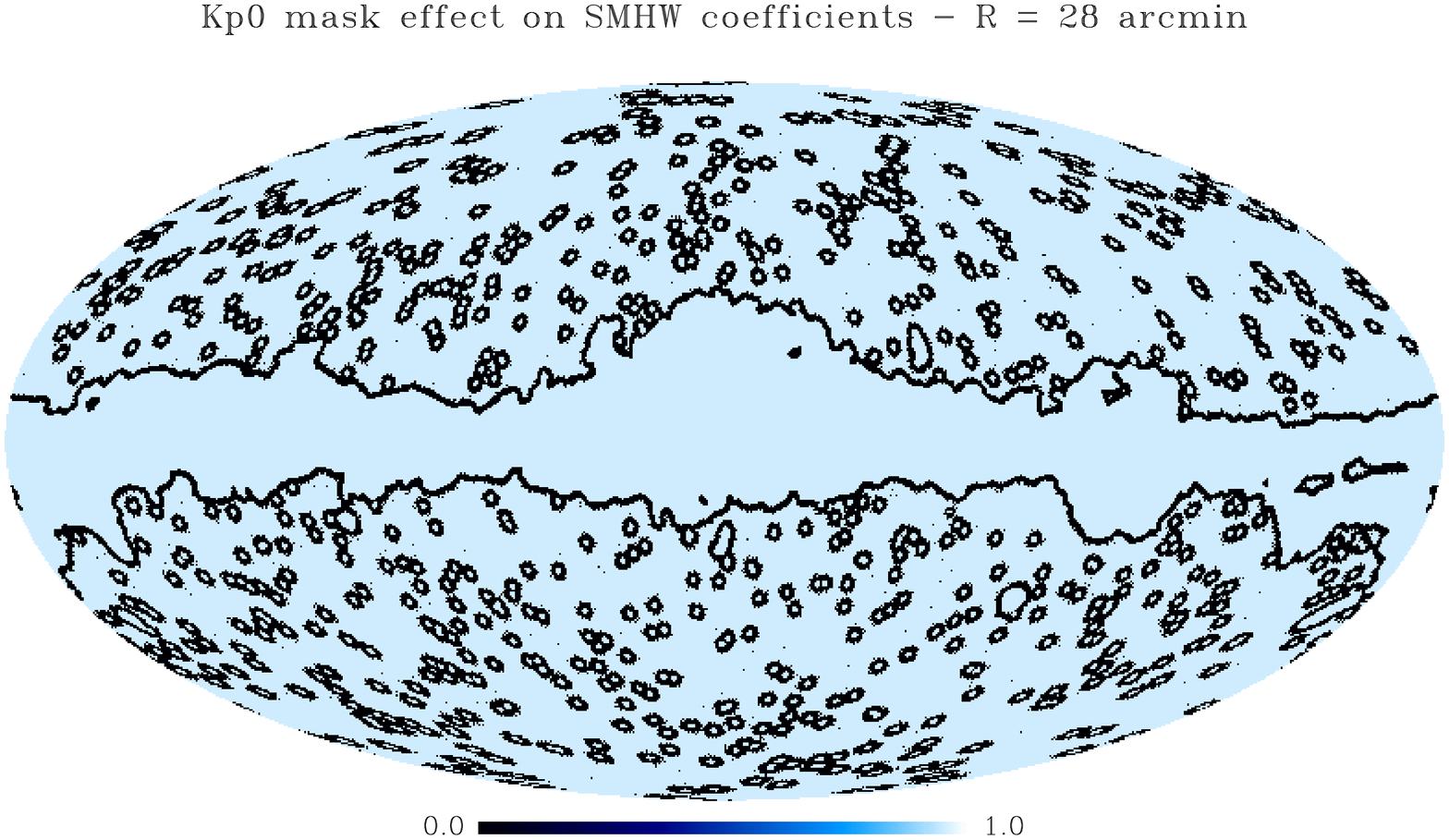}
\end{center}
\caption[nvss_{m}ask_{e}ffect]{{\protect\small {Effect of incomplete sky coverage, modeled by the WMAP Kp0 mask, visualized by plotting on a sky map the quantity defined in Eq.~\ref{discrep}. From top to bottom, the result for needlets, flat binning, and SMHW ($28^\prime$).}}}
\label{kp0_mask_effect}
\end{figure}

Due to their localization properties, needlets are especially useful in the
analysis of partial sky coverage. In fact, in view of Eq.~\ref{expine} we
expect the value of needlets coefficients to be mildly affected by the
presence of gaps in the maps. To illustrate this important feature, we plot
the quantity 
\begin{equation}
\chi _{jk}=\frac{\langle(\beta _{jk,mask}-\beta _{jk})^{2}\rangle}{\langle\beta _{jk}^{2}\rangle}
\label{discrep}
\end{equation}
in Fig.~\ref{kp0_mask_effect}, where the Kp0 mask\footnote{See LAMBDA website, http://lambda.gsfc.nasa.gov/}, that is used to remove Galactic emission and point sources from WMAP data (roughly $75\%$ of the sky), is applied. The
expected values of Eq.~\ref{discrep} are again evaluated by means of 100 Monte Carlo simulations; in particular we focus on needlets coefficients
corresponding to $B=2.72$ and $j=5$, which amounts to a range in multipoles
space in the order of $58<\ell <398$. To put our results in perspective, in the same Figure
we show analogous findings with the use of a tophat binning filter and SMHW. We remind the SMHW formula
\begin{eqnarray}
\Psi(y,R)&=&\frac{1}{\sqrt{2\pi}N(R)}\Big[1+\Big(\frac{y}{2}\Big)^2\Big]^2\times\nonumber\\
&\times&\Big[2-\Big(\frac{y}{R}\Big)^2\exp{(-y^2/2R^2)}\Big]\nonumber
\end{eqnarray} 
where $y=2\tan{\theta/2}$ ($\theta$ is the polar angle), $R$ is the scale of convolution and $N(R)$ a normalization factor \citep{2002MNRAS.336...22M}.

\begin{figure}
\begin{center}
\includegraphics[width=\columnwidth]{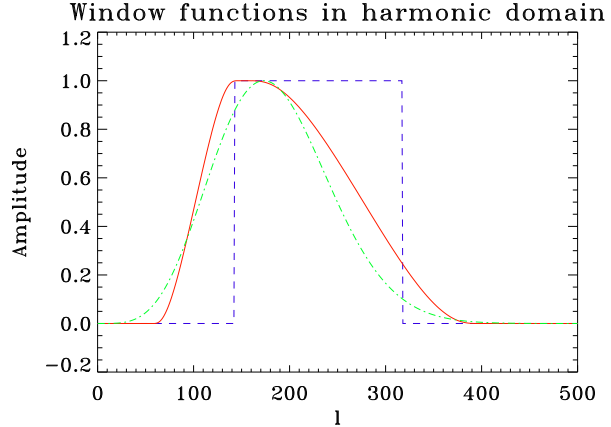}
\end{center}
\caption[filters_{s}hape]{The red solid line represents needlets window function, $b(\frac{\ell}{B^{j}})$ in harmonic space for $B=2.72$, $j=5$. The blue dashed and green dot-dashed lines provide the tophat and the SMHW window functions, respectively. The SMHW corresponds to a scale $R=28\prime$ in pixel space.}
\label{filters_shape}
\end{figure}

Under these circumstances, needlets coefficients are well
localized, but slightly sensitive to the mask. Indeed, only $56\%$ of
the pixel are changed by less than $0.1$; SMHW coefficients perform a bit better ($73\%$) while a simple tophat binning fails completely (only $6\%$). The difference between the two wavelets bases can be due to the different power that they give to multipoles (see Fig.~\ref{filters_shape}). In fact the performance of needlets can be improved choosing the appropriate $B^j$, that defines the optimal shape for the window $b(\cdot)$, given the multipoles range of interest. For details see \cite{Cardoso2007}. In the same paper, the authors argue that an optimal filter can be adapted to deconvolve a specific mask: this property provides a further degree of flexibility to the needlets toolbox. 

%

In \cite{Baldi2006},  another relevant
property of needlets coefficients was discussed, namely their asymptotic
uncorrelation at any fixed angular distance, for growing frequencies $j$. More
explicitly, at high frequency needlets coefficients can be approximated as a
sample of identically distributed and independent (under Gaussianity)
coefficients, and this property opens the way to a huge toolbox of
statistical procedures for CMB data analysis (for instance, for testing Gaussianity and isotropy). Also, in view of Eq.~\ref{needcof}, for full sky maps and in the absence of
any mask we should expect the theoretical correlation to be identically zero
whenever $\left| j_{1}-j_{2}\right| \geq 2$.

Let us define the realized correlation between two different scales $j_1$, $j_2$ as
\begin{equation}
\label{correlation_expansion}
\rho _{j_{1}j_{2}}=\frac{\sum_{k}\langle\beta _{j_{1}k}\beta _{j_{2}k}\rangle}{\sqrt{
\sum_{k}\langle\beta _{j_{1}k}^{2}\rangle\sum_{k}\langle\beta _{j_{2}k}^{2}\rangle}}\textrm{.}
\end{equation}
By using Eq.~\ref{needcof} one has that:
\begin{eqnarray*}
\langle\beta _{j_{1}k}\beta _{j_{2}k}\rangle&=&\sum_\ell b(\ell/B^{j_1})\,b(\ell/B^{j_2})\,C_\ell\sum_m K_{\ell\ell m m} \\
\end{eqnarray*}
where $C_\ell$ is the underlying CMB angular power spectrum and the coupling kernel $K_{\ell\ell^\prime m m^\prime}$ is defined in terms of the observed mask $W(\hat\gamma)$ (see e.g.\ \citealt{2002ApJ...567....2H}):
\[
K_{\ell\ell^\prime m m^\prime}=\int_{S^2}Y_{\ell m}(\hat\gamma)Y_{\ell^\prime m^\prime}(\hat\gamma)W(\hat\gamma)d\Omega
\]
Note that, in the absence of gaps (i.e.\ $W(\hat\gamma)=1$), $\sum_m K_{\ell\ell m m}$ reduces simply to $(2\ell+1)$.

Equation (\ref{correlation_expansion}) can not be expected to be reproduced exactly, due to numerical approximations; in particular, we should stress that
theoretical results are derived under the assumption that needlets coefficients are
evaluated at \emph{exact} cubature points, so that the $\left\{
a_{lm}\right\} $ are precisely reconstructed from the maps. Of course, this
is not the case in practice; however, we do expect small and vanishing
values for $j_{1}\ll j_{2}$. At the same time, we expect this correlation to
increase on the presence of sky cuts, but less so than for other bases. Here, we want to illustrate the practical
relevance of this mathematical results by means of simulations on the
correlation coefficient. More precisely, we computed the quantity (\ref{correlation_expansion}) 
by performing a Monte Carlo over $100$ simulations. Our findings are shown in Tables~\ref{corrt},~\ref{maskcorrt}.

\begin{table}
\caption{Needlets correlation parameter. $B=2.72$ without gaps}
\label{corrt}
\begin{center}
\begin{tabular}{|c|c|c|c|c|c|}
$j/j^\prime$ & $1$ & $2$ & $3$ & $4$ & $5$ \\ \hline
1 & 1.000 & 0.275 & 0.001 & 0.001 & 0.003 \\ \hline
2 & - & 1.000 & 0.248 & 0.001 & 0.001 \\ \hline
3 & - & - & 1.000 & 0.268 & 0.001 \\ \hline
4 & - & - & - & 1.000 & 0.242 \\ \hline
5 & - & - & - & - & 1.000 \\ \hline
\end{tabular}
\end{center}
\end{table}

\begin{table}
\caption{Needlets correlation parameter. $B=2.72$ with gaps}
\label{maskcorrt}
\begin{center}
\begin{tabular}{|c|c|c|c|c|c|}
$j/j^\prime$ & $1$ & $2$ & $3$ & $4$ & $5$ \\ \hline
1 & 1.000 & 0.420 & 0.140 & 0.040 & 0.060 \\ \hline
2 & - & 1.000 & 0.335 & 0.023 & 0.001 \\ \hline
3 & - & - & 1.000 & 0.291 & 0.004 \\ \hline
4 & - & - & - & 1.000 & 0.252 \\ \hline
5 & - & - & - & - & 1.000 \\ \hline
\end{tabular}
\end{center}
\end{table}

We view these results as very encouraging. In the absence of a mask, the correlation coefficient is by any practical means virtually negligible for all frequency distances greater or equal than 2, while at distance $\Delta j=1$ the correlation is around $\sim0.25$ in good agreement with eqn.~\ref{correlation_expansion} which predicts $0.22$ for our input parameters. In the presence of sky cuts, the
performance deteriorates as expected only at low $j$ where it  exceeds a few
percentage points, as shown for our simulations in the case of the Kp0 mask. A computation analogous to (\ref{correlation_expansion}) yields for SMHW the theoretical results reported in Table~\ref{theoCorr_SMHW}; note how we have non zero values at all lags. Numerical results to support the theoretical findings are provided by Tables~\ref{SMHWcorrt},~\ref{SMHWmaskcorrt}.
\begin{table}
\caption{Theoretical correlation (full sky) for needlets and SMHW}
\label{theoCorr_SMHW}
\begin{center}
\begin{tabular}{cccccc}
corr/$\Delta j$ & 0 & 1 & 2 & 3 & 4 \\\hline
Needlets & 1.000 & 0.220 & 0.000 & 0.000 & 0.000 \\ \hline
SMHW & 1.000 & 0.500 & 0.100 & 0.014 & 0.002 \\ \hline
\end{tabular}
\end{center}
\end{table}
We believe these compared results strongly support the potential of
needlets for the implementations of statistical procedures, where
uncorrelation properties are clearly a very valuable asset.

\begin{table*}
\caption{SMHW correlation parameter without gaps. The scale $R$ is given in arcmin.}
\label{SMHWcorrt}
\begin{center}
\begin{tabular}{|c|c|c|c|c|c|c|c|}
$R/R^\prime$ & 1792 & 896 & $448$ & $224$ & $112$ & $56$ & $28$ \\ \hline
1792 & 1.000 & 0.503 & 0.109 & 0.016 & 0.002 & 0.0002 & 0.00003\\ \hline
896 & - & 1.000 & 0.500 & 0.099 & 0.014 & 0.002 & 0.0002\\ \hline
448 & - & - & 1.000 & 0.510 & 0.103 & 0.014 & 0.002\\ \hline
224 & - & - & - & 1.000 & 0.511 & 0.104 & 0.014\\ \hline
112 & - & - & - & - & 1.000 & 0.513 & 0.107\\ \hline
56 & - & - & - & - & - & 1.000 & 0.519\\ \hline
28 & - & - & - & - & - & - & 1.000\\ \hline
\end{tabular}
\end{center}
\end{table*}

\begin{table*}
\caption{SMHW correlation parameter with gaps. The scale $R$ is given in arcmin.}
\label{SMHWmaskcorrt}
\begin{center}
\begin{tabular}{|c|c|c|c|c|c|c|c|}
$R/R^\prime$ & $1792$ & $896$ & $448$ & $224$ & $112$ & $56$ & $28$\\ \hline
1792 & 1.000 & 0.496 & 0.113 & 0.022 & 0.005 & 0.002 & 0.0007 \\ \hline
896 & - & 1.000 & 0.520 & 0.115 & 0.021 & 0.005 & 0.002\\ \hline
448 & - & - & 1.000 & 0.523 & 0.114 & 0.021 & 0.005 \\ \hline
224 & - & - & - & 1.000 & 0.520 & 0.114 & 0.020 \\ \hline
112 & - & - & - & - & 1.000 & 0.522 & 0.116 \\ \hline
56 & - & - & - & - & - & 1.000 & 0.527\\ \hline
28 & - & - & - & - & - & - & 1.000\\ \hline
\end{tabular}
\end{center}
\end{table*}

\begin{figure*}
\label{correlationfigs}
\includegraphics[width=\columnwidth]{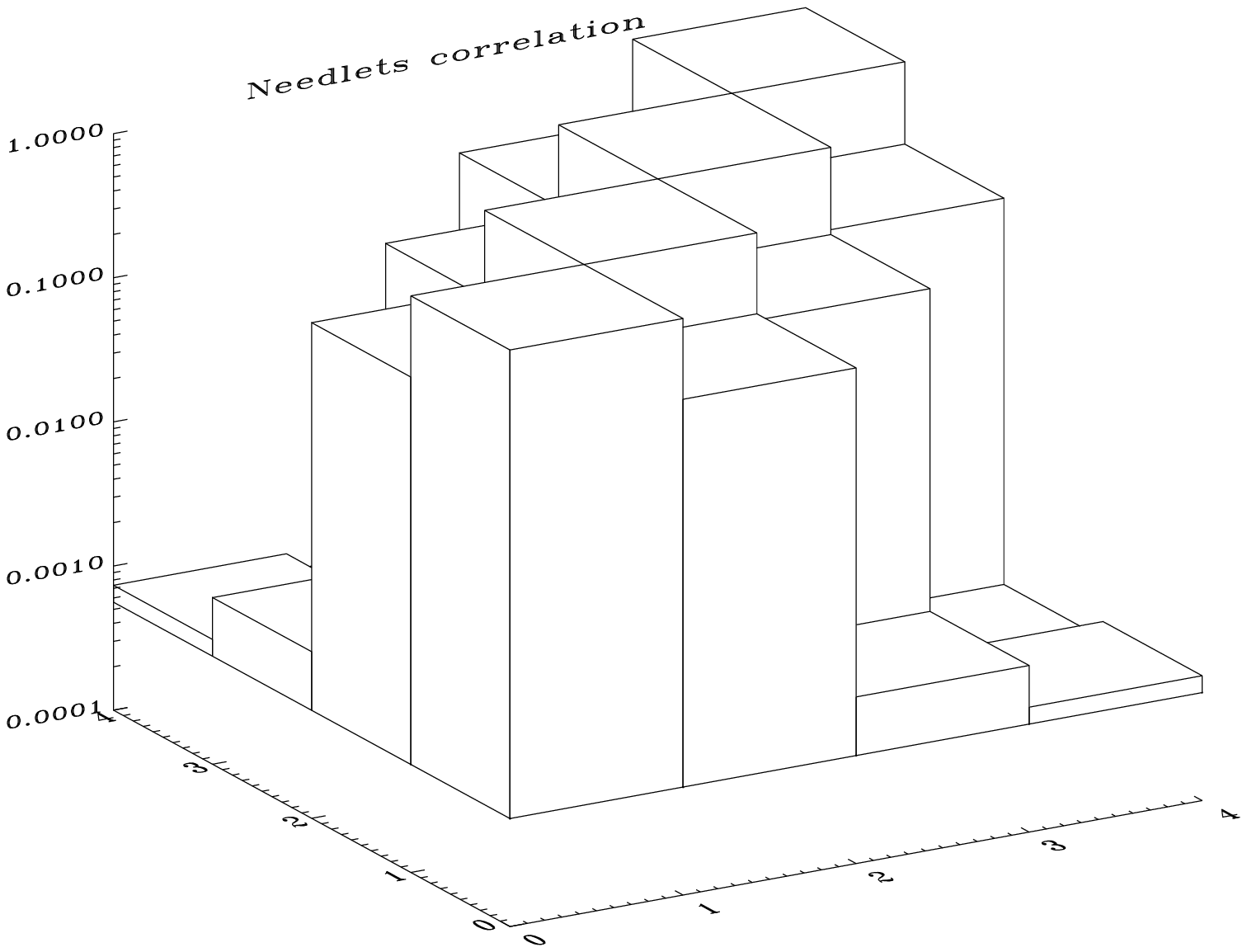}
\includegraphics[width=\columnwidth]{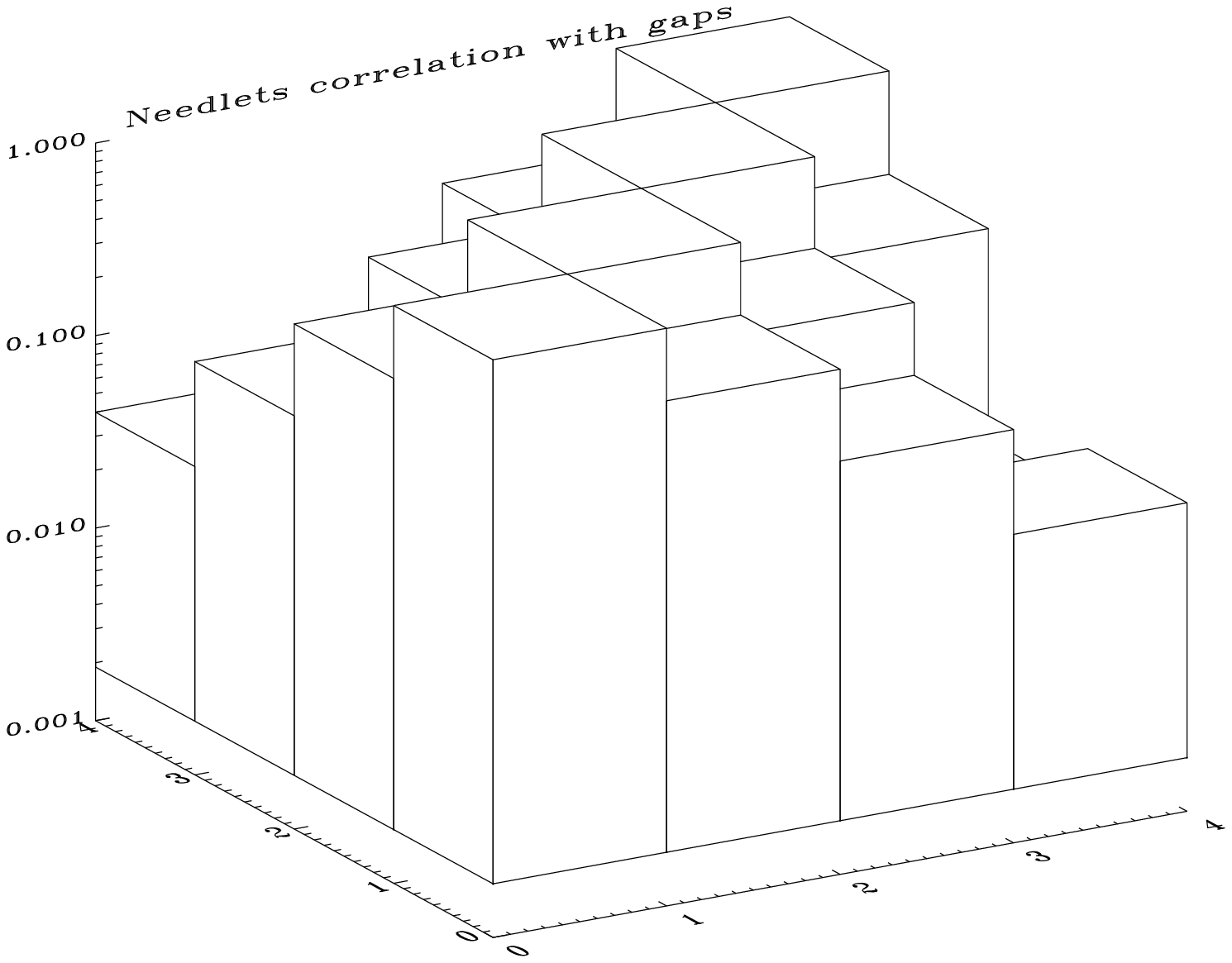}
\includegraphics[width=\columnwidth]{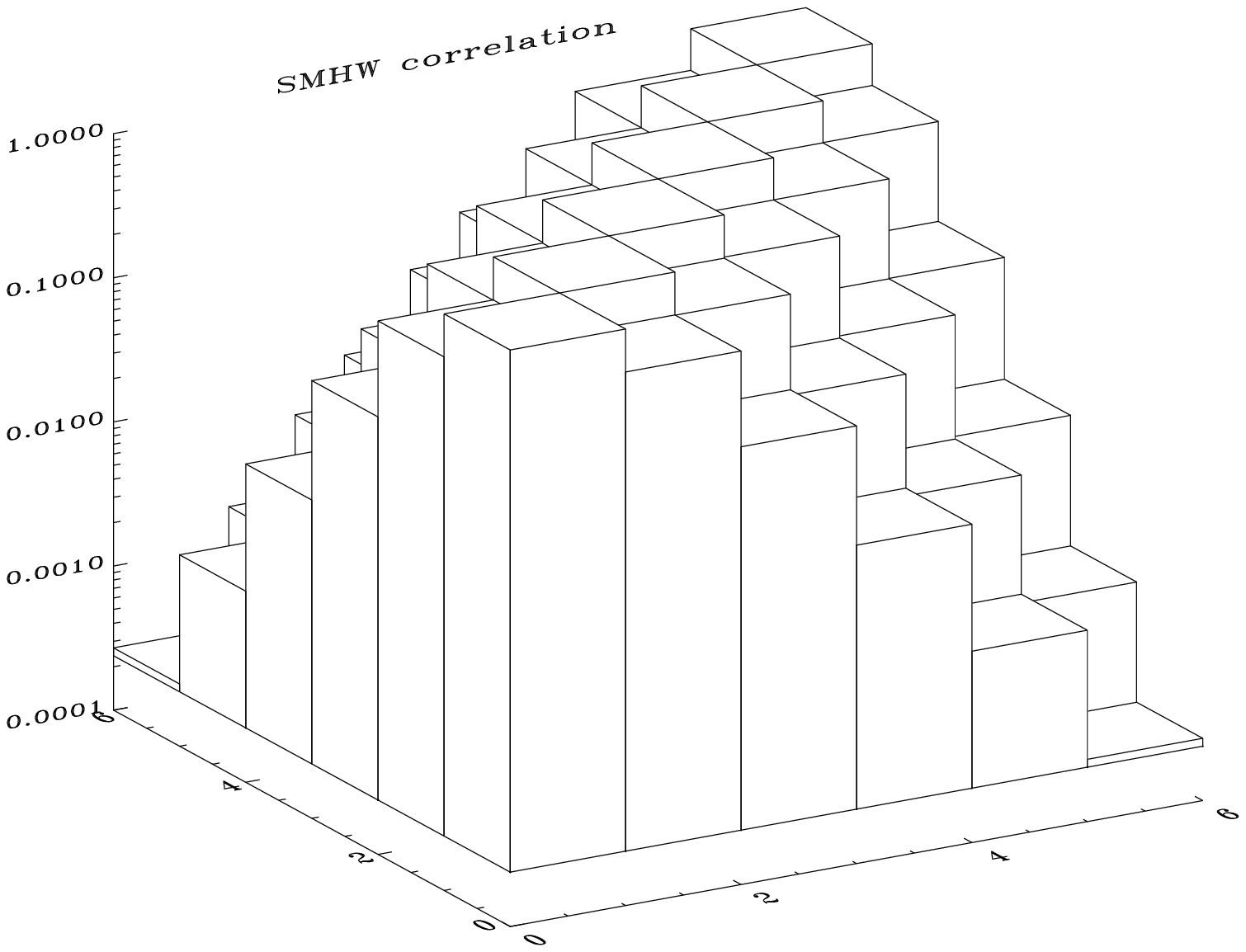}
\includegraphics[width=\columnwidth]{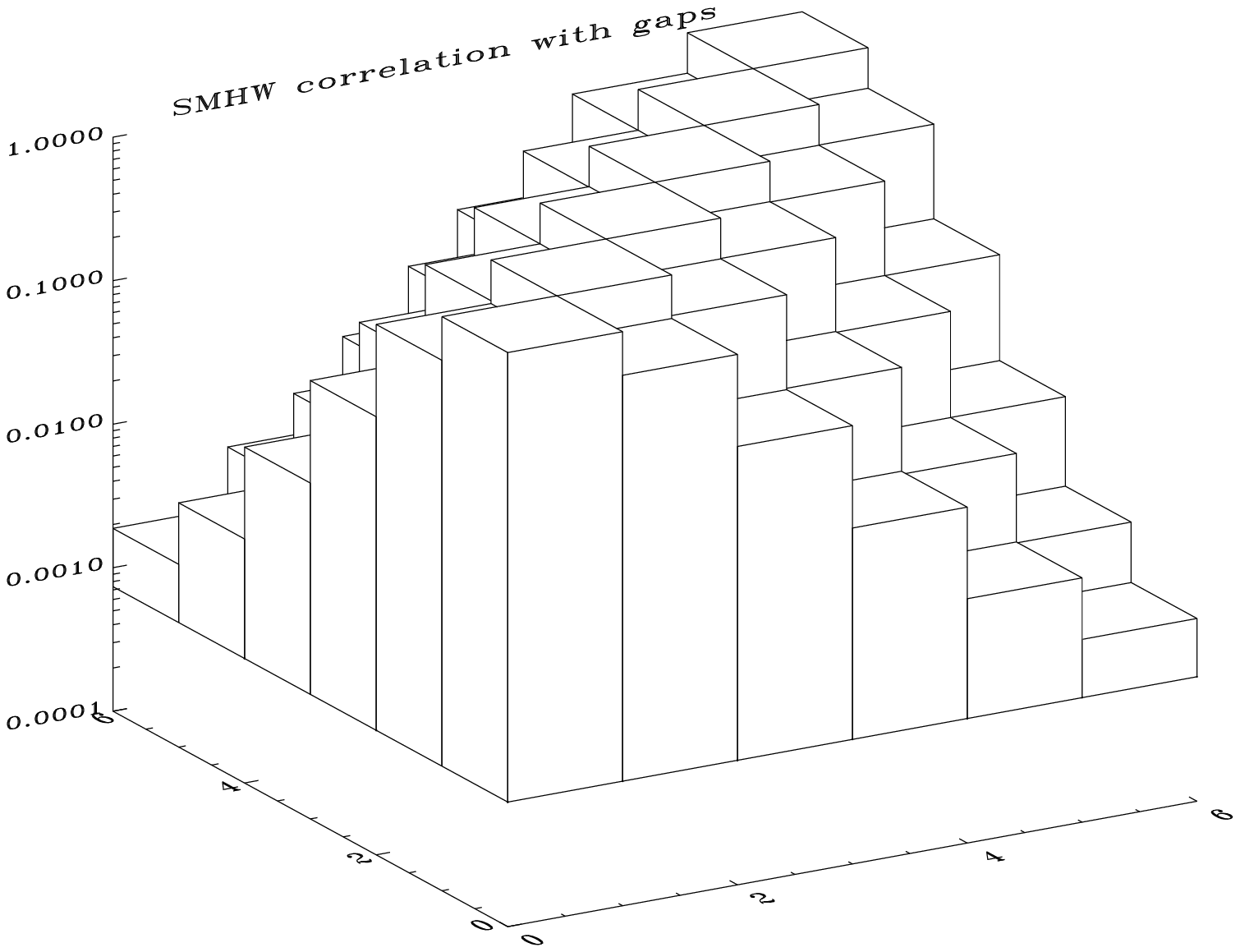}
\caption{The correlation for needlets (top panels) and SMHW (bottom panels) is plotted as a function of the scale, in absence (left panels) and in presence (right panel) of sky cuts (Kp0 mask).}
\end{figure*}

\begin{figure}
\begin{center}
\includegraphics[width=\columnwidth]{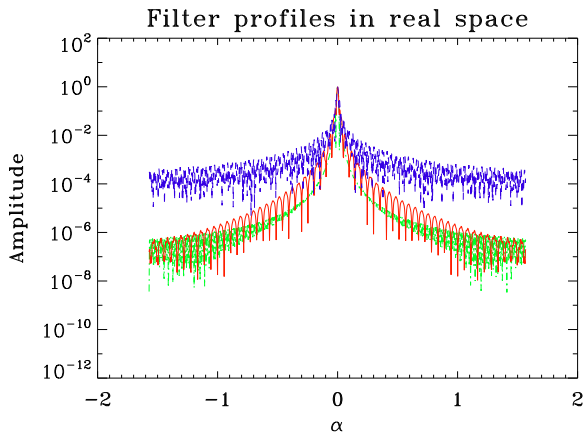}
\end{center}
\caption[localization_comparison]{Behaviour of needlets (solid red), SMHW (dot-dashed green) and tophat binning (blue dashed) in pixel space. The angle in horizontal axis is measured in radiants.}
\label{localization}
\end{figure}

\begin{figure}
\begin{center}
\includegraphics[width=\columnwidth]{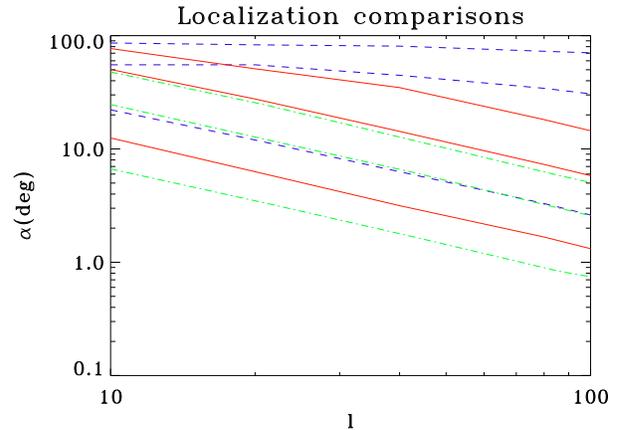}
\end{center}
\caption[localization_{c}omparison2]{The three lines represent the angle at
which the area of needlet, in red, or the tophat, in blue, filter reaches
the $99\%$, $95\%$, $68\%$ of the total area as a function of the peak multipole in each window function. The latter corresponds directely to a given $j$ for needlets and to the scale $R$ for SMHW; for the tophat window the central $\ell$ in the band is taken.}
\label{localization2}
\end{figure}

As a further comparison, we evaluated the domains in harmonic and real spaces for needlets, tophat binning and SMHW. In particular we normalized the three bases to have roughly an equal area in the harmonic domain, paying attention to have the maximum of the power in a similar range of multipoles. Results are plotted in Fig.~\ref{filters_shape},~\ref{localization}. It is evident how SMHW and needlets outperform tophat binning by two order of magnitudes in terms of localization in real space: indeed in this domain the two wavelets constructions perform quite similarly. Moreover, in Figure \ref{localization2}, we computed the angle where the integral of the filter functions in pixel space reaches $68\%$, $95\%$ and $99\%$ of the total area, respectively. Again, it is
immediate to check how at every scale needlets outperform very clearly a
simple binning approach; on the other hand, SMHW seems slightly more concentrated in this setting. The linear trend for needlets in the
log-log plot is a direct conseguence of their construction, and in
particular of the functional dependence on $\ell /B^{j}$.

On the other hand the advantage of needlets over SMHW emerges quite clearly in the harmonic domain. More precisely, after normalizing the two methods to be centred at the same angular scale, with roughly the same total power, the needlets support seems clearly more concentrated than SMHW. In particular we stress how SMHW suffer from ``leakage'' by the very low multipoles, i.e. exactly those most affected by sky cuts and cosmic variance. No such leakage occurs for needlets.
 
\begin{figure}
\begin{center}
\includegraphics[width=\columnwidth]{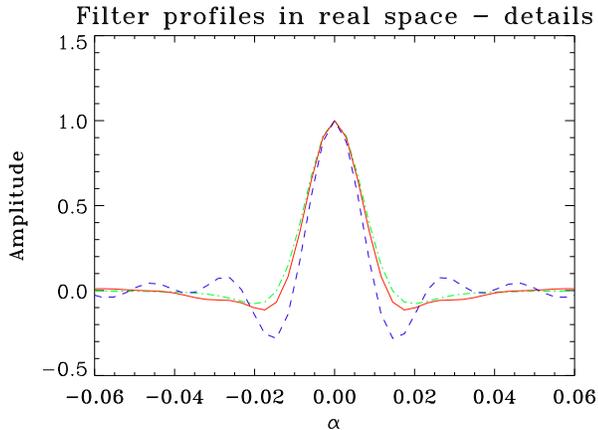}
\end{center}
\caption[details]{Here in this figure we provide details of the behaviour in pixel space over the relevant range, i.e. the region where the three functions exceed $0.001$. Lines have the same meaning as in the previous figures.}
\label{zoom}
\end{figure}

\section{Concluding remarks}\label{conclusions}

In this paper, we have described a new construction for a spherical wavelets
frame, and we have illustrated several of its properties; most notably
localization properties in the real and harmonic spaces, uncorrelations of
the resulting random coefficients, and independence from any tangent plane
approximations. Moreover, needlets enjoy a direct reconstruction property
which allows analysis and synthesis to be implemented directly and in a
computationally convenient manner. Each of these properties has been
illustrated by means of Monte Carlo simulations. The encouraging results
reported suggest that needlets can become a valuable tool in the several
areas of CMB data analysis where other wavelets have already proved useful.

\section*{Acknowledgments}
Some of the results in this paper have been derived using the HEALPix (G\'orski et al., 2005) package.

\bibliographystyle{mn}
\bibliography{biblio}

\end{document}